\documentclass[11pt]{article}
\pdfoutput=1
\usepackage{jheppub}
\usepackage{graphicx}
\usepackage{amsfonts,amssymb,amsmath}
\usepackage{mathrsfs}
\usepackage{braket}
\usepackage[hang]{subfigure}
 \usepackage[utf8]{inputenc}
\usepackage[a4paper,top=1.89in, bottom=0in, left=2in, right=0in]{geometry}
\usepackage{xcolor}

\usepackage{color}
\usepackage[none]{hyphenat}

\newcommand{\be}{\begin{equation}}
\newcommand{\ee}{\end{equation}}
\newcommand{\ba}{\begin{eqnarray}}
\newcommand{\ea}{\end{eqnarray}}

\title{Thermodynamics of de Sitter Black Holes with Conformally Coupled Scalar Fields}
\author[1,2]{Fil Simovic}
\author[2]{Danny Fusco}
\author[1,2]{Robert B. Mann}

\affiliation[1]{Perimeter Institute for Theoretical Physics,\\
	31 Caroline St. N.,Waterloo, Ontario N2L 2Y5, Canada}
\affiliation[2]{Department of Physics and Astronomy, University of Waterloo,\\
	Waterloo, Ontario N2L 3G1, Canada}
\emailAdd{fil.simovic@gmail.com}
\emailAdd{dfusco@uwaterloo.ca}
\emailAdd{rbmann@uwaterloo.ca}

\abstract{We investigate the thermodynamic properties of 3+1 dimensional black holes in asymptotically de Sitter spacetimes, conformally coupled to a real scalar field. We use a Euclidean action approach, where boundary value data is specified at a finite radius `cavity' outside the black hole, working in the extended phase space where the cosmological constant is treated as a thermodynamic pressure. We examine the phase structure of these black holes through their free energy. For the MTZ subclass of solutions, we find Hawking-Page-like phase transitions from a black hole spacetime to thermal de Sitter with a scalar field. In the more general case, Hawking-Page-like phase transitions are also present, whose existence depends further on a particular cosmic censorship bound.}


\begin{document}
\maketitle

\section{Introduction}

For some 40 years, Hawking's discovery that black holes radiate as blackbodies \cite{hawking1975}, and the resulting elevation of the laws of black hole mechanics \cite{bekenstein1973,bardeen1973} to true thermodynamic laws, has guided research in general relativity, quantum gravity, and quantum information, due in large part to the questions these observations raise about the unitarity of black hole evaporation, and the nature of those regimes where general relativistic and quantum effects are on equal footing. Though many open questions remain, undoubtedly the study of the thermodynamic properties of black holes has elucidated many aspects of not only quantum field theory in curved spacetime, but also string theory, information theory, and the microscopic structure of gravity itself. At the core of these ideas is the universality of the Bekenstein-Hawking entropy, which has been calculated using a half-dozen or so completely disparate approaches \cite{jacobson1994b,youm1999,ashtekar1998,carlip2000,solodukhin2011,bianchi2013}, each of which provide a unique perspective on the state-counting interpretation of $S=A/4$.

Perhaps one of the most striking features of black holes arising from their thermodynamic properties is that they, like ordinary thermodynamic systems, can undergo phase transitions. This phenomenon, first discovered by Hawking and Page for anti-de Sitter (AdS) black holes \cite{hawking1983}, has since been shown to occur generically in a wide variety of black hole spacetimes, and in theories beyond Einstein-Hilbert gravity \cite{kubiznak2016,belhaj2012,gibbons2005c,hennigar2017h,frassino2014}. These phase transitions have a natural interpretation in the context of AdS/CFT \cite{maldacena1998}, where they are dual to a deconfinement transition in the boundary CFT \cite{witten1998a}, providing not only a way of studying strongly coupled systems nonperturbatively, but also giving insights into the unitarity of black hole evaporation and the information problem\footnote{The transition between constant modes of the gauge field and the free particle state for the boundary CFT is explicitly unitary. This transition maps to the evaporation of the black hole in the bulk thus explicitly demonstrating the unitarity of black hole evaporation, though of course this is not a complete resolution to the problem as the intermediate state is not easily calculated \cite{harlow2016}.}.

Recent investigations into the phase structure of black holes have largely taken place in the extended phase space, where the cosmological constant is treated as a thermodynamic variable that acts like a pressure/tension. This follows from observations by Kastor, Ray, and Traschen that when variations of $\Lambda$ are included in the derivation of the first law and Smarr relation, a new potential appears with the interpretation of an effective thermodynamic volume \cite{kastor2009}. This extended phase space is the natural context in which to study the phase structure of black hole spacetimes with non-zero $\Lambda$, where already many new thermodynamic phenomena have been discovered, such as triple points, re-entrant phase transitions, superfluid transitions, and more \cite{altamirano2014b,frassino2014,hennigar2017f}. These phenomena are not only interesting in their own right, drawing parallels between some of the most exotic objects that appear in our universe and everyday thermodynamic processes like the liquid-gas phase transition of ordinary water, but also have implications for gauge theories through the AdS/CFT correspondence, each example having a non-trivial dual description in terms of the boundary CFT \cite{karch2015,sinamuli2017,wen2007}. This extended phase space, which we consider here, is the natural context in which to study asymptotically dS/AdS black holes\footnote{A review of extended phase space thermodynamics can be found in \cite{kubiznak2017}.}. 

Whereas a wealth of interesting phenomena in AdS spacetimes have been discovered thus far, the domain of asymptotically de Sitter (dS) spacetimes remains relatively unexplored, despite recent measurements of the cosmological constant indicating that our universe has $\Lambda>0$ \cite{planckcollaboration2019}. A number of difficulties arise in the dS case, which have historically impeded progress. For one, there is no globally timelike Killing vector field with which to associate the mass, rendering the construction of conserved charges difficult. Asymptotically de Sitter spacetimes also contain a cosmological horizon, which in general radiates at a temperature different from the black hole, thus forcing the system out of equilibrium. 

A number of approaches exist that attempt to circumvent these problems, each with their own issues and limitations. One is the effective temperature approach, where a single temperature is assigned to the entire spacetime, which depends on both the cosmological and event horizon temperatures \cite{urano2009,zhangli-chun2010}. Another approach considers subsets of the parameter space where the two horizon temperatures are equal, thus restoring a straightforward notion of equilibrium temperature for the spacetime, at the cost of severely limiting the number of situations that can be explored \cite{mbarek2019}. In this paper we instead use a Euclidean path integral approach, developed first by Gibbons and Hawking \cite{gibbons1977} and extended by York \cite{york1986,braden1990} to allow for the definition of the canonical (and later, grand canonical) ensemble, an approach which has seen renewed interest in recent years \cite{peca1999,carlip2003}. Here, boundary value data is fixed on a finite radius `cavity' surrounding the black hole, and thermodynamic quantities are determined through the fundamental relationship between the partition function and Euclidean path integral. Fixing the temperature at the cavity allows the black hole to achieve equilibrium, permitting the study of equilibrium thermodynamics. Upon taking the extended thermodynamic phase space into account, new phenomena emerge for charged de Sitter black holes, such as swallowtubes and analogues of the familiar Hawking-Page (HP), small-large, and reentrant phase transitions exhibited by AdS black holes \cite{simovic2019a,simovic2019,haroon2020a}.

In this paper, we add another layer of complexity to the puzzle. Though there is a rich history in the interplay between scalar fields and the gravitational interaction, there has been comparatively little interest in the thermodynamic aspects of black holes endowed with scalar fields, despite the strong motivations available from cosmology, field theory, and particle physics. Among these are the recent discovery of a fundamental scalar in nature (the Higgs boson), as well as the ubiquity of scalar fields in cosmology, where they are used to model dark matter, neutron stars, the inflaton field, and more \cite{ferreira1998,tsujikawa2003,liebling2017,arapoglu2019}. Indeed, scalar fields arise quite generally in effective field theory descriptions of fundamental fields after coarse-graining, so understanding their role in black hole thermodynamics is essential if we are to have a complete picture of the thermodynamic properties of astrophysical black holes. 
 
Some attempts at understanding this interplay have already been made, notably with the Bronnikov-Melnikov-Bocharova-Bekenstein (BMBB), the Martinez-Troncoso-Zanelli (MTZ) and the Achucarro-Gregory-Kuijken (AGK) pierced black hole  solutions \cite{nbocharova1970,bekenstein1974b,martinez2003,achucarro1995}. These early examples, originally the only known exceptions to the ``black holes have no (scalar) hair" theorem \cite{mayo1996} have now given way to a large number of solutions of Einstein gravity and its extensions \cite{Herdeiro:2015waa}.  However in the context of black hole thermodynamics, particularly for asymptotically de Sitter black holes, they have received relatively little attention. The BMBB black hole has been shown to lack any sensible thermodynamic interpretation \cite{Zaslavskii_2002}, owing to the divergence of the scalar field at the horizon, while also being generically unstable \cite{winstanley2004}. 
However the situation for the MTZ black hole, which is the analogous solution for $\Lambda>0$, is somewhat more promising. There, the entropy and temperature are both found to be finite, and there appears to exist non-trivial phase structure when charge is present \cite{radu2005}. However, these preliminary investigations (the ones with $\Lambda\neq0$ in particular) are not formulated in the extended phase space as we do here.
 
This leaves a number of open questions concerning the interplay between scalar fields and black hole thermodynamics. Can sensible thermodynamics be formulated with conformal coupling, or are more complicated couplings required? Do general features seen in the phase structure of de Sitter black holes (Hawking-Page transitions, swallowtubes, etc.) survive when a scalar field is introduced, or do new features appear? In order to shed some light on these questions, we consider a more general class of solutions recently reported by Anabalon and Cisterna  \cite{anabalon2012}. There, a large class of analytic (asymptotically de Sitter) solutions to Einstein gravity with a conformally coupled scalar field were discovered, a subclass of which are the well-known MTZ and BMBB solutions. In this paper we will examine the thermodynamic aspects of these Anabalon-Cisterna (AC) black holes, in an effort to answer some of the questions posed above, and towards a better understanding of black hole thermodynamics with matter fields at play in de Sitter space.

Our paper is organized as follows. In Section \ref{sec2}, we discuss the application of the Euclidean path integral approach to de Sitter spacetimes. In Section \ref{sec3} we describe the scalar theory being considered, defining the relevant quantities and commenting on various limits being considered. In Section \ref{sec4}, we calculate the on-shell Euclidean action for these conformally coupled de Sitter black holes, and determine the free energy of the spacetime. In Section \ref{sec5}, we examine the phase structure of these black holes, using the temperature to determine regions where interesting phase structure may be present, and demonstrating the presence of Hawking-Page-like transitions through the free energy. Finally, we summarize the main results and comment on future avenues of research. 

\section{The Euclidean Action Approach to Thermodynamics}\label{sec2}

The Euclidean action approach, which we employ here, provides a tool for studying gravitational thermodynamics when the spacetime under consideration does not admit a straightforward definition of temperature, as in those without global timelike Killing vector fields. Underlying the approach is the fundamental relationship between the free energy $F$, the path integral, and the partition function $\mathcal{Z}$ for a quantum field,
\be\label{free}
F=-T\log \mathcal{Z}\approx T I_E\ ,
\ee
where $I_E$ is the Euclidean action of the classical spacetime. The second equality holds in the semi-classical approximation, and comes directly from the path integral over metrics of Euclidean signature ($t\rightarrow i\tau$) which are $\beta$-periodic in imaginary time and with fixed boundary geometry,
\be
\mathcal{Z}=\int\hspace{-0.1cm}\mathcal{D}[g]\ e^{-I_E[g]/\hbar}\ .
\ee
To leading order in $\tfrac{m_p}{M}$, matter fields do not contribute to the path integral, allowing one to sum only over geometries. This quantity can further be approximated by considering only the dominant contribution to the integral, which comes from metrics that are classical solutions to the equations of motion, namely those for which $\delta I_{E}[g_{cl}]=0$. This is the saddle point approximation, and gives that
\be
\int\mathcal{D}[g]\ e^{-I_E[g]/\hbar}\approx e^{-I_{E}[g_{cl}]/\hbar}\ ,
\ee
where we denote by $g_{cl}$ said metrics. In the canonical ensemble, which we consider here, the thermodynamic potential related to the Euclidean action corresponds to the Helmholtz free energy $F$. From the above considerations, we have that
\be
\mathcal{Z}\approx e^{-I_{E}[g_{cl}]/\hbar}\qquad\rightarrow\qquad F\approx TI_E\ .
\ee
In this approximation, the partition function is further related to other thermodynamic quantities through the usual formulas from statistical mechanics:
\be\label{eands}
E=\frac{\partial I_{E}}{\partial \beta}\ ,\qquad S=\beta\frac{\partial I_{E}}{\partial \beta}-I_{E} \ .
\ee
The variations above are done with all other parameters held fixed, possibly necessitating the inclusion of extra terms to account for the dependency of $\beta$ on those parameters\footnote{This is necessary if, for example, $\beta=\beta(r_+)$, since this imposes a constraint that does not allow the action to vary with respect to $\beta$ while keeping $r_+$ fixed.}. This kind of approach has been used to study thermodynamic aspects of successively more exotic spacetimes (particularly with de Sitter asymptotics), such as neutral and charged de Sitter black holes \cite{carlip2003,simovic2019a}, Born-Infeld black holes \cite{simovic2019}, and Gauss-Bonnet black holes \cite{haroon2020a}, demonstrating a wide variety of thermodynamic phenomena.

\section{Self-Interacting Scalar Fields}\label{sec3}

In light of the motivations discussed above, we will consider the AC class of exact solutions to Einstein's equations with non-zero cosmological constant and a conformally coupled, self- interacting scalar field \cite{anabalon2012}. This large class of solutions   describes a variety of topologically distinct spacetimes, including everywhere-regular black holes and wormholes, which further have everywhere regular scalar field configurations. The general theory is described by the following action $I$, potential $V(\phi)$, and equation of motion for $\phi$,
\be\label{action1}
I=\int\!\!\! \sqrt{-g}\left[\dfrac{R-2\Lambda}{16\pi}-\frac{1}{2}(\partial\phi)^2-\frac{1}{12}\phi^2R-V(\phi)\right]d^4x\ ,
\ee
\be
V(\phi)=\alpha_1\phi+\alpha_3\phi^3+\alpha_4\phi^4\ , \qquad \Box\phi=\frac{1}{6}R\phi+\frac{\partial V}{\partial \phi}\ ,\nonumber
\ee
\be
\frac{1}{8\pi}(G_{\mu \nu}+\Lambda g_{\mu\nu})=\partial_{\mu} \phi \partial_{\nu} \phi-\frac{1}{2} g_{\mu \nu}(\partial \phi)^{2}-g_{\mu \nu} V(\phi)+\frac{1}{6}\left(g_{\mu \nu} \square-\nabla_{\mu} \nabla_{\nu}+G_{\mu \nu}\right) \phi^{2}\ ,\nonumber
\ee
where we have set $G=1$, and the cosmological constant determines the de Sitter length scale through $\Lambda=3/l^2$. The scalar field here is conformally coupled to gravity, circumventing early `no-hair' theorems formulated either in asymptotically flat spacetimes or with minimally coupled scalar fields (for reviews see \cite{bekenstein1995,herdeiro2018}). In any case, conformal coupling is the unique prescription for which the Green's functions of the field reduce locally to their Minkowski space counterparts, and propagation of massive particles on the light cone is avoided, making the above theory an interesting candidate towards understanding the effects of coupling matter fields to black hole spacetimes.

With the choices $\alpha_{1}=-\frac{3}{4\pi}\,\alpha_{3}$, $\alpha_{3}=-\Lambda\sqrt{\frac{16\pi}{27}} \frac{\xi}{\xi^{2}+1}$, and $\alpha_{4}=-\frac{2\pi \Lambda}{9}$, the equations of motion have an exact  solution given by
\be
\phi(r)=\left(\frac{3}{4\pi}\right)^{\!\!\!1/2}\frac{M(\xi-1)-\xi r}{M(\xi-1)+r}\ ,
\ee
while the Euclidean\footnote{The Lorentzian counterpart is easily obtained via Wick rotation $\tau\to i t$.} metric for the spacetime is
\be\label{metric}
ds^2=\Omega(r)\!\left[f(r)\,d\tau^2+\frac{dr^2}{f(r)}+r^2d\Omega_2^2\,\right]
\ee
\be
\Omega(r)=\frac{(r+(\xi-1)M)^2}{(r-M)^2}\ ,\qquad f(r)=\left(1-\frac{M}{r}\right)^{\!\!2}\!-\Sigma^2 r^2\ ,\qquad \Sigma\equiv\sqrt{\frac{\,\Lambda\,(\xi^2-1)^2}{3\,(\xi^2+1)}} \nonumber\ .
\ee
Here, $\xi$ is an arbitrary dimensionless parameter yielding the MTZ solution when equal to zero, $M$ is the mass parameter, $\Lambda$ is the cosmological constant, and $d\Omega_2^2$ is the metric on $\mathcal{S}^2$. We have also defined $\Sigma=\Sigma(\Lambda)$ for convenience. Though a number of geometrically distinct solutions arise in various limits of the general theory defined above (detailed in \cite{anabalon2009}), we focus mainly on the case where $\Lambda>0$ and $0\leq\xi<1$. This gives rise to a black hole spacetime with the conformal structure of the Schwarzschild-de Sitter solution, with a new asymptotic region replacing the inner singularity, and an everywhere finite value for the scalar field (as long as $r>M$).

The focus on positive $\xi$ arises from the fact that the spacetime possesses a curvature singularity at $r=M(1-\xi)$ due to the divergence of the scalar field there. When $\xi\geq 0$, this singularity is always hidden behind the event horizon, while for $\xi<0$ one can have a naked singularity for certain values of $\{M,\xi,\Lambda\}$. There is also a curvature singularity at $r=0$. Note that the conformal factor $\Omega$ diverges on the horizon in the limit $M\rightarrow0$, unless the limit $\xi\rightarrow0$ is taken first.

The spacetime possesses three distinct horizons: a cosmological, an event, and an inner horizon. The cosmological and event horizons are  located at
\be\label{horizons}
r_h=\frac{1-\sqrt{1-\mathstrut 4 M \Sigma}}{2 \Sigma}\ ,\qquad r_{\text{cosmo}}=\frac{1+\sqrt{1-\mathstrut 4 M \Sigma}}{2 \Sigma}\ .
\ee
The Nariai limit, where the event horizon and cosmological horizon coincide, is  achieved if
\be
M_{\text{Nariai}}=\frac{1}{4\Sigma}=\frac {\sqrt {3 \left( {\xi}^{2}+1 \right)}  }{ 4\sqrt {\Lambda}\left( {\xi}^{2}-
		1 \right) }\ .
\ee
In the Nariai limit, the only possible location for the cavity is at $r_c=r_{\text{Nariai}}$, where both the entropy and energy (as determined by \eqref{eands}) diverge. However they diverge in such a way that the temperature acquires a finite value.

\section{Calculating the Euclidean Action}\label{sec4}

In this section, we evaluate the on-shell Euclidean action for the theory defined above. We refer the reader to \cite{braden1990} for a more detailed discussion of the procedure, covering only the essential points here. The action \eqref{action1} is supplemented by a Gibbons-Hawking-York-type boundary term, required for the action to be stationary and to have the correct composition properties for the path integral \cite{gibbons1977,dyer2009a}. In the case of the scalar-tensor theory above, the required boundary term produces the following Euclidean action
\be\label{action2}
I_E=\int\!\!\! \sqrt{g}\left[\dfrac{\Lambda}{8\pi}+V(\phi)-\dfrac{1}{4}\Box\phi^2\right]d^4x-2\!\int\!\! \sqrt{k}\left[\frac{1}{16\pi}+\frac{1}{12}\phi^2\right]\!K\ d^3\!x\ ,
\ee
where the equations of motion have been used to simplify the bulk part. In the boundary part, $k_{ab}$ is the metric on the boundary (with determinant $k$), and $K$ is the trace of the extrinsic curvature of the boundary defined by an inward pointing normal vector $s^a$ as $K=\tfrac{1}{2}\mathcal{L}_s{k_a}^a$. The bulk part of the action is integrated from $r_h$ to $r_c$, while the boundary part is evaluated at $r=r_c$. As usual, the requirement that the $\mathcal{S}^2$ corresponding to the event horizon be non-degenerate further requires that the periodicity in $\tau$ be fixed to
\be
\beta_h=\frac{4\pi}{\ f'(r)\big|_{r_h}}=\frac{2\pi}{\Sigma\,\big(1-2\,\Sigma\,r_h\big)}\ .
\ee
This is the inverse Hawking temperature of the black hole (and is invariant under conformal rescaling of the metric), which reduces to the known temperature of the MTZ solution when $\xi=0$. Equilibrium in the system is achieved by fixing the temperature at the cavity $\beta_c^{\,-1}$ to simply be the Hawking temperature redshifted to the cavity radius, thus determining the periodicity in locally observed time for quantum fields at $r_c$, namely
\be\label{betac1}
\beta_c=\beta_h\sqrt{f(r_c)\,\Omega(r_c)}\ .
\ee
The contributions to the Euclidean action \eqref{action2} are
\begin{align}
V(\phi)&=\frac {\Lambda\xi}{\sqrt {3\pi} \left( \xi^2+1 \right) }\phi-\frac{\Lambda\!\sqrt{16\pi}}{27}\frac {\xi}{\xi^2+1}\phi^{3}-\frac{2\pi\Lambda}{9}\phi^4\ ,
\\
\Box\phi^2&=\frac{4f\phi\,\phi'}{\Omega}\left(\frac{1}{r}+{\Omega'}{\Omega}\right)+\left(\frac{2f\phi\,\phi'}{\Omega}\right)^2\ ,\\
K&={\frac {r\,\Omega f'+3\,rf   \Omega' +4\,\Omega f  }{ \Omega^{2}\,r}\sqrt {{\frac {\Omega  }{f  }}}}\ ,
\end{align}
where primes indicate derivatives with respect to $r$. Performing the integrations gives the on-shell Euclidean action for the black hole spacetime
\begin{align}\label{actionb}
I_B=&\dfrac{\beta_c\,r_c}{6}\left(1- \xi^2\right)\!  \Big[ 
\Lambda\left( r_c-r_h \right)  \big(  \left( r_c+r_h \right) \Sigma-1 \big)  \Big( {\Sigma}^{2}r_c^{4}+
\left( 3\,\Sigma\,\Xi+1 \right)  \left( \Xi+r_c \right)\left( \Sigma\,r_c^{2}+\Xi+r_c \right)  \Big) \nonumber\\
&\times  \big(r_c+\Xi(1-\xi)\big)  \left( \xi^2-1 \right) ^{2}-3\,{\Sigma}^{3}\left( \xi^2+1 \right)  \Big( 2\,r_c^{5}-3\,r_c^{7}\,{\Sigma}^{2}+{\Xi}^{5} \left( \xi-1 \right)+r_c\,{\Xi}^{4} \left( 2+9\,\xi \right)\nonumber\\
&  +{\Xi}^{3}r_c^{2} \left( 14+3\,r_c^{2}\,{\Sigma}^{2} \left( \xi-1 \right) +19\,\xi \right) +\Xi\, r_c^{4}\left( 11+4\,\xi-3\,r_c^{2}\,{\Sigma}^{2} \left( \xi+4 \right) \right) \nonumber\\
&+{\Xi}^{2}r_c^{3} \left(20+15\,\xi -3\,r_c^{2}{\Sigma}^{2}
\left( \xi+4 \right)  \right)  \Big)  \Big]
\end{align}
where $\Xi\equiv r_h (\Sigma\, r_h-1)$. In the semi-classical approximation, Equation \eqref{free} is justified by observing that the dominant contributions to the path integral come from metrics that are classical solutions to the equations of motion. In our case, there are two distinct ones\footnote{Note that Minkowski space is not a solution to the classical equations of motion when $\Lambda\neq0$, and so it does not contribute to the path integral. This justifies the use of the empty de Sitter space as the ``background subtraction'' instead of the commonly employed flat space.}: the Schwarzschild-de Sitter-like solution given by \eqref{metric}, and the ``empty'' de Sitter-like solution where $M=0$. The path integral over \eqref{metric} contains contributions from graviton fluctuations far away from the black hole. In order to isolate the contribution to the partition function coming from the black hole itself, we subtract from the black hole action the same action evaluated for $M=0$, matching the metrics on the boundary $\partial\mathcal{M}$. In this way, we are effectively setting the reference point for the free energy to be the empty de Sitter spacetime. We will often refer to this phase as the `radiation phase' or `thermal de Sitter'. Note that even when $M=0$, the scalar field does not vanish, though it does achieve a constant value. 

This choice of background subtraction is not so important when dealing with, for example, charged black holes in the canonical ensemble, because the pure radiation phase is anyways inaccessible to black holes with fixed non-zero charge. However, in our case it is critical, since a shift in the free energy above/below the $F=0$ line (which corresponds to the radiation phase) may wash out any Hawking-Page-like phase transitions present. The Euclidean action for the reference spacetime is
\\
\be
I_0=\frac {\beta_c\,r_c\Big[\!\left( 1-{\xi}^{4} \right) \left( 9\,{\Sigma}^{2}r_c^{2}-6 \right) -\Lambda\,r_c^{2} \left( \xi^2-1 \right) ^{3} \Big]  \big( r_c+r_h \left( \Sigma\,r_h-1 \right)  \big) }{6\big( \xi^2+1\big)\sqrt{\mathstrut 1-{\Sigma}^{2}r_c^2\,}\,\big[ r_c-r_h \left( \Sigma\,r_h-1 \right)  \left( \xi-1 \right)  \big] }\ .
\ee
With these considerations the Euclidean action for the Schwarzschild-de Sitter black hole with conformally coupled scalar field is simply
\be\label{action3}
I_E=I_B-I_0\ .
\ee
This is in contrast to the Euclidean action obtained in \cite{barlow2005}, where instead of the canonical ensemble considered here, a microcanonical ensemble without boundary is used for the MTZ black hole. One can obtain this result by taking the limit $r_c\rightarrow r_{\text{cosmo}}$ in \eqref{actionb}, which gives
\be
I_E=0\ .
\ee
The action in this case identically vanishes as there is no charge for the scalar field, in agreement with Equation (22) of \cite{barlow2005}. Our result is more general though, since we have not set $\xi=0$. Even in this more general case the action vanishes without charge when no cavity is present. The energy and entropy of the spacetime can then be easily determined through \eqref{eands}, though we omit the results here due to their lengthiness.

\section{Thermodynamics and Phase Structure}\label{sec5}

The action \eqref{action3} allows us to calculate all thermodynamic quantities of interest  through the canonical relations \eqref{eands}. In particular, we will be interested in examining the equilibrium temperature \eqref{betac1} and free energy \eqref{free} to look for possible phase transitions within the spacetime. In the following analysis, we work in units where $\hbar=C=G=1$.

\subsection{The First Law}

Before analyzing the phase structure, we give a brief discussion of the first law of thermodynamics for these black holes. The first law, as formulated in the extended phase space for uncharged, asymptotically dS/AdS black holes reads
\be\label{firstlaw}
dE=TdS+VdP+\sigma dA\ .
\ee
Here, $E$ and $S$ are the energy and entropy as determined by \eqref{eands}, $T$ is the temperature, $V$ is the thermodynamic volume, and $P$ is the pressure, related to the cosmological constant by
\be
P=-\frac{\Lambda}{8\pi}=\frac{(d-1)(d-2)}{16\pi G}
\ee
for a spacetime of dimension $d$ \cite{kubiznak2017}. Appearing also is a work term $\sigma dA$ due to the presence of the cavity, where $A$ is the geometric area of the boundary and $\sigma$ is interpreted as the surface tension. The conjugate variables $V$ and $\sigma$ are determined by enforcing \eqref{firstlaw} as a constraint. The expressions are lengthy and offer little insight so we omit them here, noting that it is simple to check that the first law is satisfied for these black holes, and that the thermodynamic volume, though in general not equal to the geometric `volume' of the spacetime, is everywhere positive.

\subsection{Equilibrium Temperature}

Working in the regime of equilibrium thermodynamics requires the assignment of a single equilibrium temperature to the system under consideration. The cavity approach, which we employ here, is particularly useful in de Sitter spacetimes as the event and cosmological horizons are generally at different temperatures, so that the system is naturally out of equilibrium. Fixing the temperature at the surface of the cavity circumvents this by forcing the spacetime to equilibrate to the cavity temperature. The spacetime considered here, in contrast to what is typical in de Sitter, is unique in that the event and cosmological horizons are actually at the same temperature, namely,
\be
T_\Delta=\frac{\ f'(r)\big|_{r_\Delta}}{4\pi}=\frac{\Sigma\,(1-2\,\Sigma\,r_\Delta)}{2\pi}\ ,
\ee 
where $\Delta=
\{h,\text{cosmo}\}$. However, this is not the temperature that an observer between the two horizons (at $r=r_c$ for example) would experience, who instead sees the black hole radiating at a redshifted temperature of 
\be
T_{c}=\frac{T_{h}}{\sqrt{\Omega(r_c)f(r_c)}}\ .
\ee
Thus, even though equilibrium in terms of particle flux is already manifest between the two horizons (as the surface gravities are equal), equilibrium as understood by observers in the spacetime is still achieved by fixing the temperature at the cavity to be equal to $T_{c}$. In this case, the temperature at the cavity required for equilibrium is
\be\label{temp}
T_c={\frac {\Sigma\, \left( 1-2\,\Sigma\,r_h\right) \big( r_c-r_h\, \left( 1-\Sigma\,r_h \right) \big) }{2\pi\big( r_c+r_h\, \left( 1-\Sigma\,r_h\right)  \left( \xi-1 \right)  \big)\sqrt { \left( \Sigma\,r_h^{2}+{\it rc}-r_h \right) ^{2}-{\Sigma}^{2}{{\it rc}}^{4}}}}\ .
\ee
With this in mind, we begin with a study of the equilibrium temperature \eqref{temp} of the spacetime under consideration. A necessary (but not sufficient) condition for the existence of complex phase structure is the multivalued nature of the horizon radius $r_h$ as a function of temperature. This is because the existence of multiple black hole phases at a given temperature is manifest in the multiple (real, positive) values of the horizon radius $r_h$, which correspond to the different sizes of black holes that exist at that temperature for a given set of fixed parameters $\{\Lambda,\xi\}$. The horizon radius thus plays the role of the order parameter for the system. These phases may be unphysical for other reasons, or otherwise be inaccessible through quasistatic transformations, so ultimately the free energy $F$ must be evaluated to determine whether a phase transition actually occurs at the given temperature, but an examination of $T(r_h)$ nonetheless provides useful insight as to where in parameter space such transitions might exist. For example, one can immediately rule out the presence of small-large phase transitions if the temperature is a monotonic function of the horizon radius. Note that even if $T(r_h)$ is monotonic, Hawking-Page-like phase transitions may still occur.
\\

\begin{figure}[h]
	\includegraphics[width=0.49\textwidth]{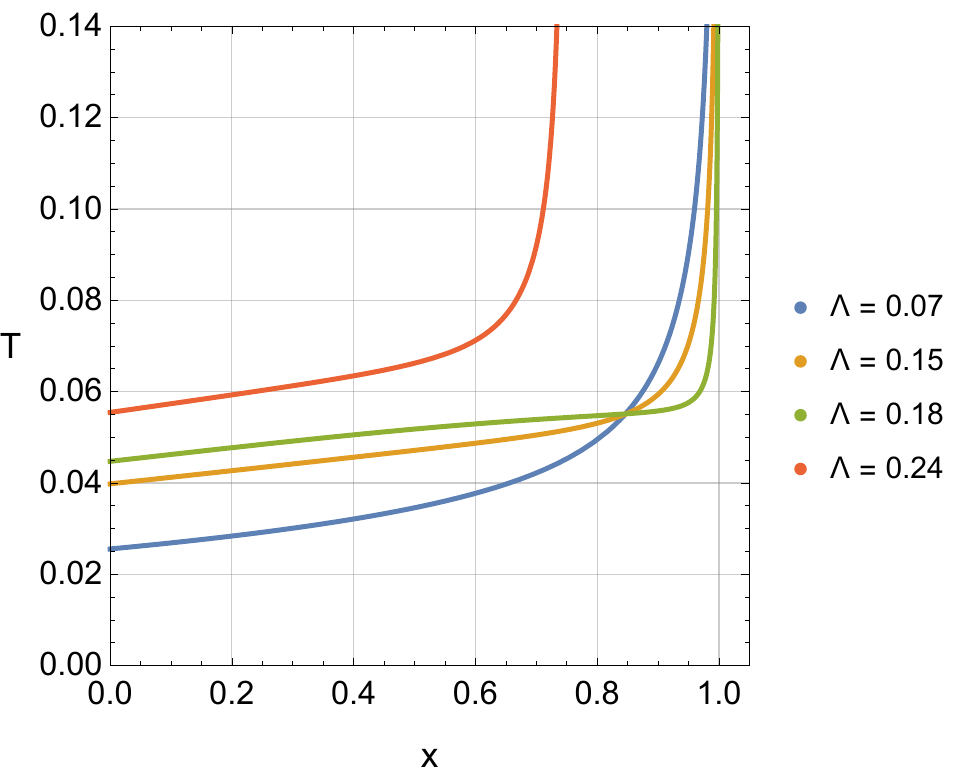}\quad\includegraphics[width=0.49\textwidth]{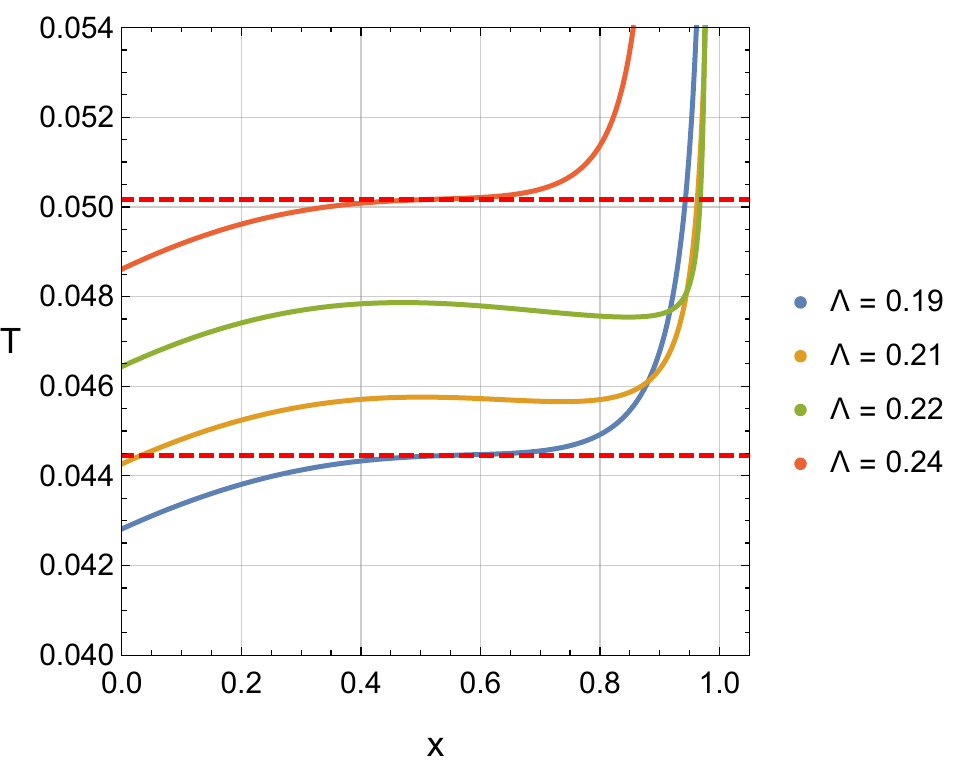}
	\caption{Equilibrium temperature $T$ as a function of $x=r_+/r_c$ for fixed cavity radius $r_c=2$, showing regions where $r_h$ is multivalued at fixed temperature, signaling a possible phase transition. \textbf{Left:} Varying pressure $\Lambda$ with $\xi=0$. \textbf{Right:} Varying pressure $\Lambda$ with $\xi=0.2$.}\label{figure1}
\end{figure}

In Figure \ref{figure1}, we plot the equilibrium temperature $T(r_h)$ of the black hole spacetime  for various choices of $\Lambda$ and  $\xi$. For convenience, the function is written in terms of the relative size of the black hole with respect to the cavity, $x\equiv r_h/r_c$. On the left of Figure \ref{figure1}, we consider different values of $\Lambda$ with $\xi=0$, corresponding to the MTZ black hole.  We observe that the temperature is always positive and monotonically increasing with respect to the size of the black hole\footnote{The apparent crossings near $x=0.8$ are coincidental.}, eventually diverging as the black hole fills the cavity entirely ($r_h\rightarrow r_c$). For large enough $\Lambda$ (as in the red curve) the temperature asymptotes before $r_h$ reaches $r_c$. This corresponds to a black hole whose mass is large enough to pull the cosmological horizon inside the cavity for the given value of $\Lambda$. Since we require $r_h<r_c<r_{\text{cosmo}}$, this places an upper limit on the black hole size. Despite the apparent jump in the figure, the new maximum moves smoothly from $x=1$ as $\Lambda$ increases. When $\Lambda<\Lambda_{\text{max}}$, the cosmological horizon always lies outside of the cavity regardless of the size of the black hole. This maximal value, below which the entire spacetime is available to the black hole, occurs at
\be\label{lambdamax}
\Lambda_{\text{max}}=\frac{3 \left(\xi^2+1\right)}{4\,r_c^2\big(\xi^2-1\big)^{\!2}}\ .
\ee
In the case of the MTZ black hole, we therefore expect to only see a Hawking-Page-like phase transition, which we will confirm by examining the free energy.

The case for non-zero $\xi$ is markedly different. On the right of Figure \ref{figure1}, we again plot $T(r_h)$ for varying $\Lambda$, this time with $\xi=0.2$. In this case, there is a region bounded by a minimum and maximum pressure (indicated by the red horizontal lines) where $r_h$ is multivalued for fixed $T$. The three distinct values of $r_h$ that exist at a particular equilibrium temperature in this region represent (up to) three coexistent black hole phases. Further analysis is required to assess whether these phases are accessible to the system. For example, the divergence of the scalar field $\phi(r)$ and the vanishing of the conformal factor $\Omega(r)$ (which leads to a singular inverse metric) place further constraints on the valid parameter space. Both of these occur at $r=M(1-\xi)$, requiring that $r_h>M(1-\xi)$. From \eqref{horizons} this places a constraint on $\xi$ such that
\be\label{constraint}
(1-\Sigma\, r_h)(1-\xi) < 1
\ee
for non-zero $r_h$ in order for the divergence to be hidden behind the event horizon. This `cosmic censorship bound' is always satisfied for $\xi\geq 0$, while for $\xi<0$ it can be violated.

\subsection{MTZ Black Holes $(\xi=0)$}

We consider first the case where $\xi=0$, corresponding to the de Sitter MTZ black hole \cite{martinez2003}. In this limit, the metric and scalar field simplify in the following ways:
\be
\phi(r)=\sqrt{\frac{3}{\!4\pi}}\,\frac{M}{r-M}\ ,\qquad V(\phi)=\alpha_4\phi^4
\ee
\be
\Omega(r)=1\ ,\qquad f(r)=\left(1-\frac{M}{r}\right)^{\!\!2}\!-\frac{\Lambda r^2}{3}\ .
\ee
The spacetime has the geometry of the lukewarm Reissner-Nordstrom-de Sitter black hole \cite{romans1992,mann1995}
 \begin{figure}[h]
	\centering
	\includegraphics[width=0.7\textwidth]{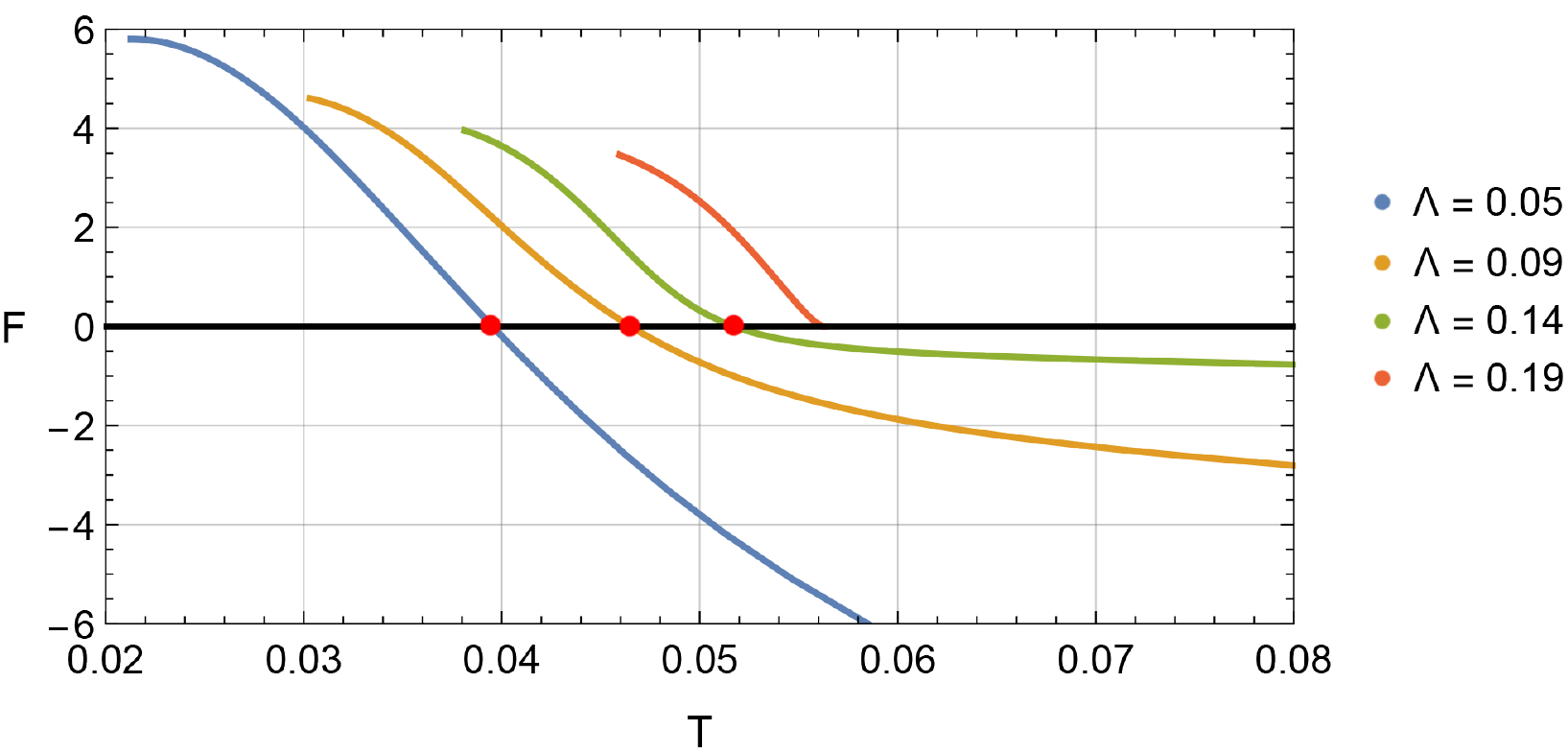}
	\caption{Free energy of the MTZ ($\xi=0$) black hole in the canonical ensemble, with varying $\Lambda$. 
The black $F=0$ line corresponds to the radiation phase. Red dots mark the critical temperature at which the black hole begins to dominate the thermal ensemble and a first-order Hawking-Page-like phase transition occurs.}\label{figure2}
\end{figure}

In Figure \ref{figure2}, we plot the free energy $F(T)$ of the MTZ black hole as a function of the cavity temperature (parametrically using $r_h$). The black line, where $F=0$, corresponds to the free energy of the empty\footnote{Empty in the sense there is no black hole. There is still a scalar field present in the spacetime.} de Sitter spacetime (the radiation phase). For each colored curve, the black hole size increases from left to right, with the leftmost end terminating where $M=0$. For low temperature, the radiation phase globally minimizes the free energy. As the temperature increases, eventually a critical point is reached (marked by a red dot) where the free energy of the black hole becomes lower than that of the radiation phase, and a Hawking-Page-like first-order phase transition occurs. As the black hole size increases further, the free energy monotonically decreases until the cavity radius is reached, where the temperature diverges. When $\Lambda$ becomes large enough (as in the red curve, where $\Lambda=\Lambda_{\text{max}}$) the Nariai limit is achieved at exactly the cavity radius, so that the line terminates on the right side at $r_h=r_c=r_{\text{cosmo}}$. 
 
Note that the presence of Hawking-Page-like transitions depends to some extent on the choice of normalization of $F$. For example, if instead of using the action of the empty spacetime as the `subtraction' term, as was done in \eqref{action3}, one simply takes the limit of \eqref{actionb} as $r_h\rightarrow0$ to determine the zero point for the free energy, then the total action never changes sign and no transition occurs. If, however, one uses Minkowski space as the reference spacetime, the curves of Figure \ref{figure2} are shifted slightly, moving the location of the critical points, but the character of the transitions remains qualitatively the same.

\subsection{AC Black Holes $(\xi\neq 0)$}

\begin{figure}[h]
	\centering
	\includegraphics[width=0.7\textwidth]{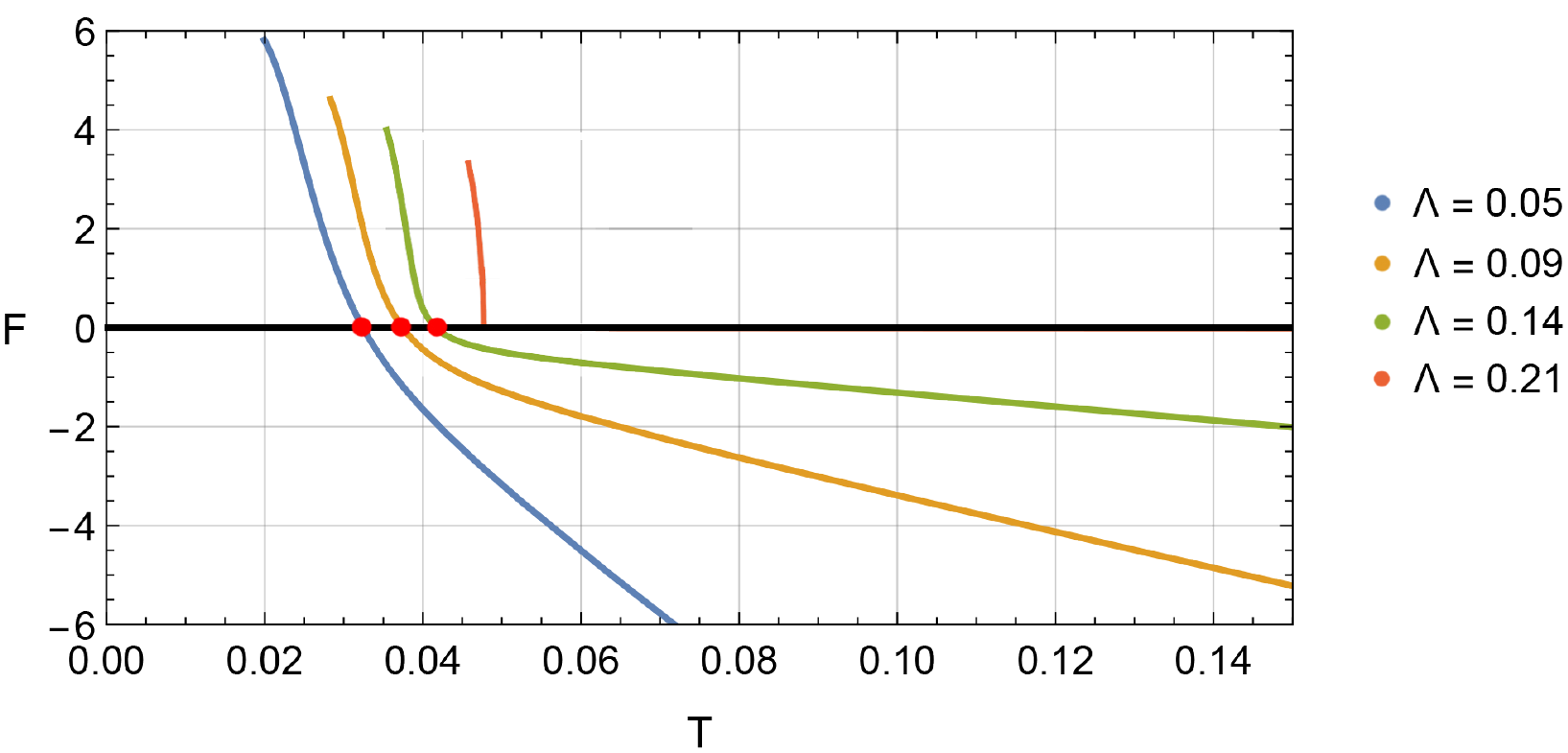}
	\caption{Free energy of the AC ($\xi\neq0$) black hole in the canonical ensemble, with varying $\Lambda$. The black $F=0$ line corresponds to the radiation phase. Red dots mark the critical temperature at which the black hole begins to dominate the thermal ensemble and a first-order Hawking-Page-like phase transition occurs.}\label{figure3}
\end{figure}

Next we consider the more general class of solutions given by $\xi\neq 0$. In this case the scalar field and metric functions are given by \eqref{metric}, with the total Euclidean action still being \eqref{action2}. In Figure \ref{figure3} we plot $F(T)$ as before, with $\xi=0.2$ and various choices of $\Lambda$. The results are qualitatively similar to that of the MTZ black hole, with Hawking-Page-like phase transitions from radiation to a large black hole occurring at a critical temperature  that depends on the cavity size and cosmological constant. As before, the black hole size $r_h$ increases from left to right along each curve, terminating at the left where $r_h=0$. There is again a maximal value for $\Lambda$ given by \eqref{lambdamax} where the cavity radius coincides with the Nariai limit, as in the red curve of Figure \ref{figure3}. Below this limit, we always have $r_h\leq r_c<r_{\text{cosmo}}$, and the temperature diverges as the black hole horizon approaches the cavity. At exactly $\Lambda=\Lambda_{\text{max}}$, the system exists at a finite temperature whose free energy is equal to that of radiation, however both the energy and entropy diverge. Despite Figure \ref{figure1} suggesting otherwise, there are no small-large black hole phase transitions present, even though for some choices of $\Lambda$ and $\xi$, $T(r_h)$ has three real roots. In all cases where this is true, only one of the roots turns out to satisfy \eqref{constraint}.

\begin{figure}[h]
	\includegraphics[width=0.49\textwidth]{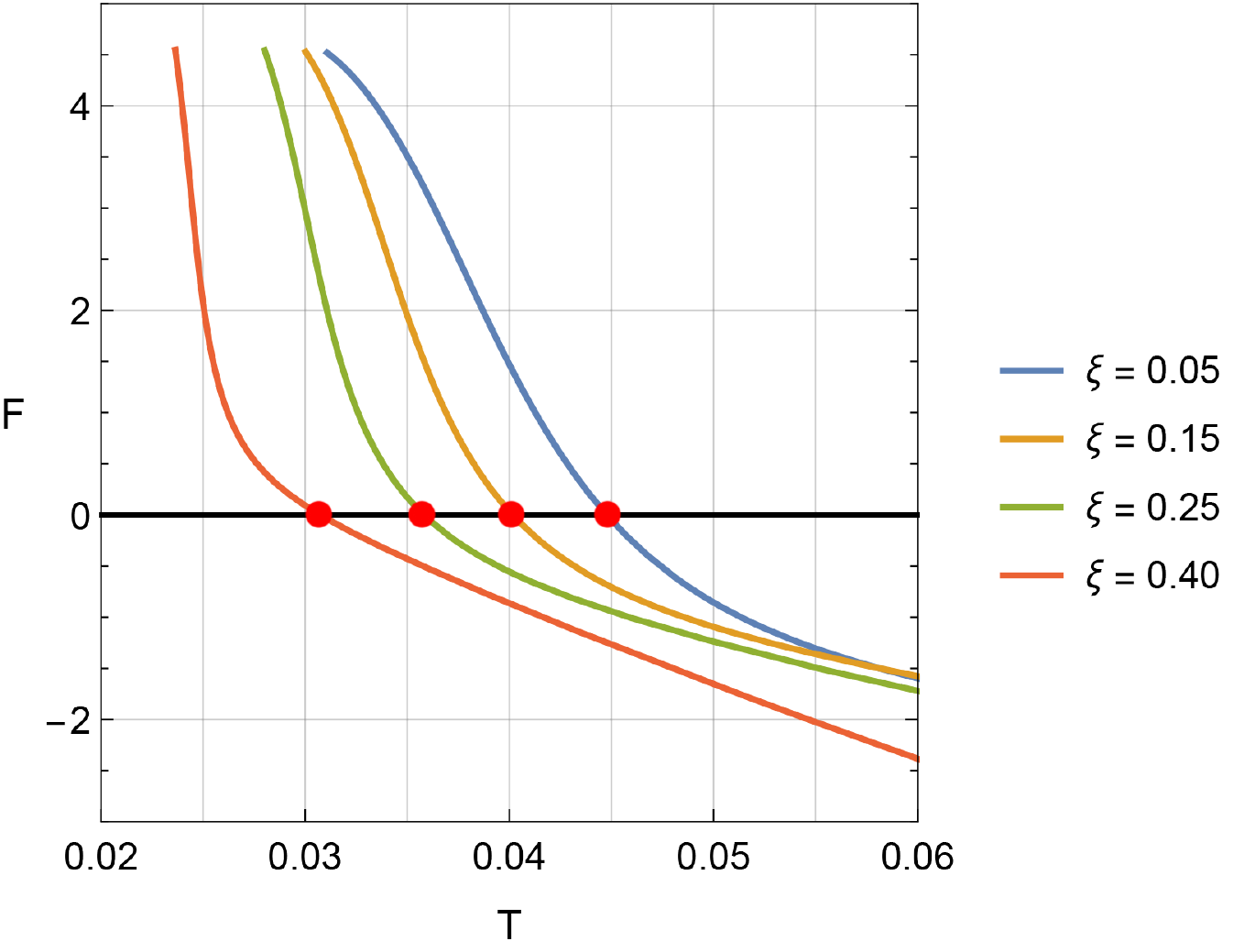}\quad\includegraphics[width=0.49\textwidth]{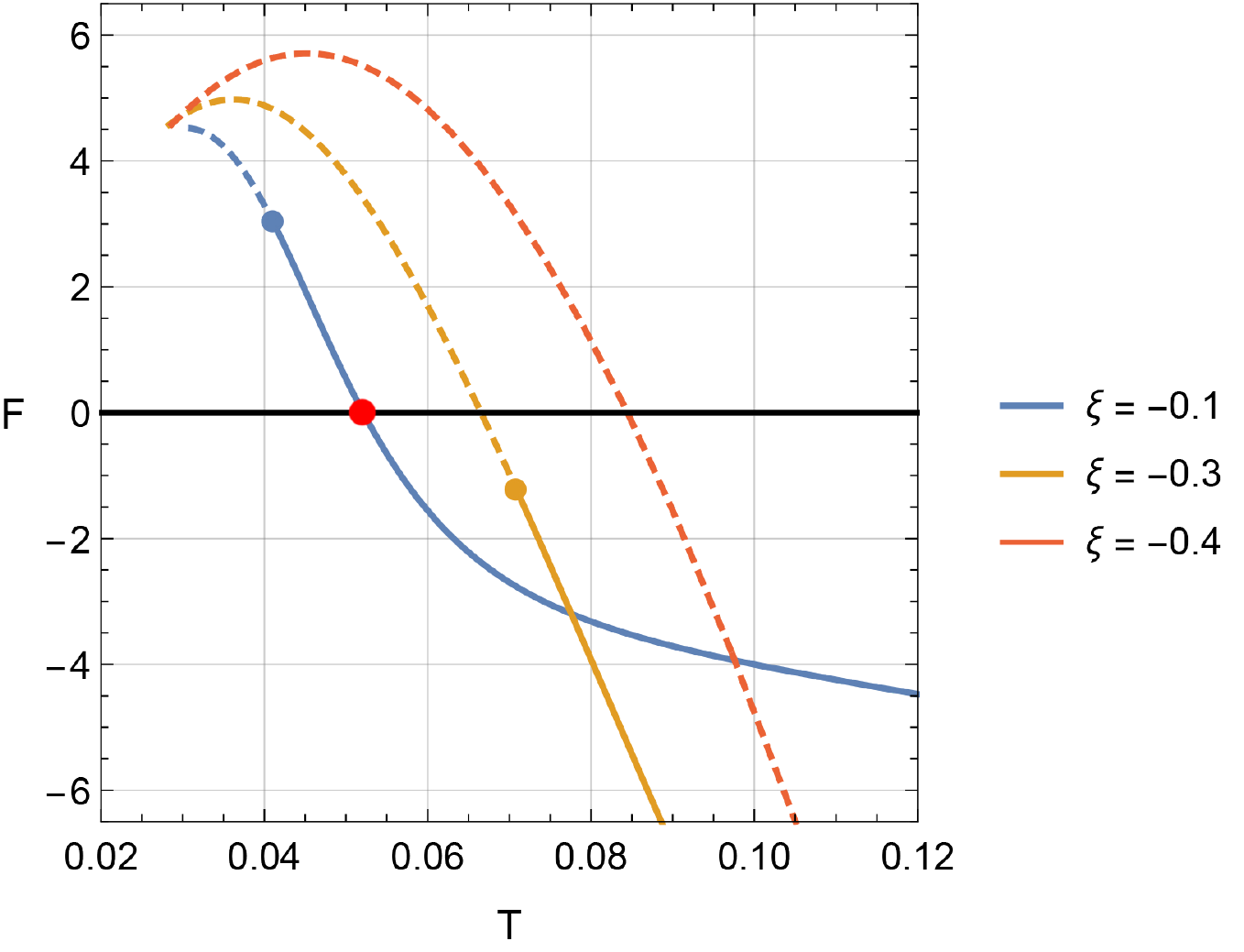}
	\caption{Free energy $F$ as a function of temperature $T$ for fixed cavity radius $r_c=2$ and varying $\xi$. For each curve, a Hawking-Page-like transition occurs at the critical temperature marked by a red point. \textbf{Left:} $\xi>0$. \textbf{Right:} $\xi<0$. Along the dashed portions of each curve, a naked singularity is present. The colored point on each curve indicates that equality in \eqref{constraint} has been reached.}\label{figure4}
\end{figure}

In Figure \ref{figure4} we again plot the free energy, this time for various $\xi$ at fixed $\Lambda$. On the left, we show positive values of the scalar parameter, while on the right we show negative values. When $\xi>0$, the behaviour is again similar to the MTZ case, where a Hawking-Page-like phase transition occurs at a critical temperature whose exact value depends on the value of $\xi$ and $\Lambda$. Again there is a minimum temperature for the spacetime, achieved when $M=0$, and the temperature diverges as the cavity is approached. 

On the other hand when $\xi<0$, qualitatively different behaviour emerges. In this case it is possible for \eqref{constraint} to be violated, so even though the free energy of the black hole crosses the $F=0$ line as it would for a Hawking-Page-like phase transition, the transition itself does not necessarily occur. On the right of Figure \ref{figure4} we plot the free energy for various negative values of $\xi$. Marked with a circle on each curve is the minimal value of $r_h$ that satisfies \eqref{constraint}, below which there is a naked singularity. The dashed portions of the curves indicate that the bound has been violated. When $\xi$ is sufficiently negative, no black hole of any size can exist within the cavity, as in the red curve where $\xi=-0.4$. For intermediate values (as in the yellow curve where $\xi=-0.2$) there is a branch of black holes that are regular outside the horizon. However, as the temperature decreases, a naked singularity forms (the Kretschmann scalar diverges) before the black hole can reach a free energy that is lower than radiation, so no transition occurs. The blue curve demonstrates a scenario where $\xi$ is sufficiently small so that the critical temperature is reached before a naked singularity forms, and a Hawking-Page transition occurs. This is the case for a given (fixed) cavity radius. It is always possible to {\it choose} the location of the cavity to ensure that the HP transition occurs before a naked singulary forms. In other words, for a given value of $\Lambda$ and $\xi$, one can always choose an appropriate value for $r_c$ such that a curve similar to the blue one is obtained, as opposed to the yellow one. In this way, the correct choice of boundary location is essential for the avoidance of naked singularities and the persistence of the HP transition.

\section{Conclusions}

We have studied the phase structure of a new class of asymptotically de Sitter black holes, conformally coupled to a real scalar field. The presence of an isothermal cavity, equivalent to fixing boundary value data on a finite surface in the spacetime, allows us to establish a notion of thermodynamic equilibrium in these asymptotically de Sitter spacetimes, which normally are not in equilibrium due to the two horizons present. What we have shown is that the Hawking-Page-like transition from a black hole spacetime to one filled with radiation occurs generically for these black holes. In the limit where the scalar field parameter $\xi$ tends to zero, corresponding to the MTZ black hole, the HP transition is present throughout most of the parameter space, except in the special case where the event and cosmological horizons simultaneously asymptote to the boundary at $r_c$ when the Nariai limit is approached. In this case, the black hole phase is unstable and always has a higher free energy than the empty spacetime.

When $\xi<0$, the situation is more complex. There is now a cosmic censorship bound that must be satisfied due to the divergence of the scalar field $\phi$ and vanishing of the conformal transformation $\Omega$. This means that while HP transitions are possible, $\xi$ must be sufficiently small for the given values of $\Lambda$ and $r_c$ in order for the phase transition to occur before a naked singularity forms in the spacetime.

Of note is the distinct absence of the `swallowtube' behaviour seen in previous examples of asymptotically de Sitter black holes. Indeed, even the typical small-large black hole phase transition does not occur for this class of black holes. This is not surprising insofar as the geometry here is that of the lukewarm Reissner-Nordstrom-de Sitter black hole \cite{romans1992,mann1995}, which represents a measure-zero element of the parameter space for those black holes. It is possible that the consideration of a charged scalar field would produce some of these more exotic transitions that are common to AdS black holes. What \textit{is} significant is the fact that the Hawking-Page phase transition appears to persist even when the geometry is coupled to matter, in this case a scalar field. Though the class of solutions considered here represents a special choice of potential for the scalar field for which exact solutions exist, we expect that similar results will hold for more `realistic' potentials. Eventually, we would like to be able to understand the thermodynamic properties of astrophysical black holes, which certainly couple to gauge fields of all kinds. This work is a step towards this direction, with the hope that we will eventually have a more complete understanding of the thermodynamic nature of the types of black holes that we see in the sky around us.

\section{Acknowledgements} This work was supported in part by the Natural Sciences and Engineering Research Council of Canada.

\bibliographystyle{JHEP} 
\bibliography{LBIB}

\end{document}